\documentclass[twocolumn,10pt,prb,aps,showpacs,superscriptaddress,preprintnumbers,amsmath,amssymb]{revtex4-1}

\usepackage{graphicx}
\usepackage{dcolumn}
\usepackage{bm}
\usepackage{float}
\usepackage[]{color}

\usepackage{tabularx}
\newcolumntype{L}[1]{>{\raggedright\arraybackslash}p{#1}}
\newcolumntype{C}[1]{>{\centering\arraybackslash}p{#1}}
\newcolumntype{R}[1]{>{\raggedleft\arraybackslash}p{#1}}

\begin{document}

\title{Tuning the hybridization and magnetic ground state of electron and hole doped CeOs$_2$Al$_{10}$: An x-ray spectroscopy study}

\author{Kai~Chen}
 \altaffiliation{Present address: Synchrotron SOLEIL, L'Orme des Merisiers, Saint-Aubin, BP 48, 91192 Gif-sur-Yvette C$\acute{e}$dex, France}
  \affiliation{Institute of Physics II, University of Cologne, Z{\"u}lpicher Stra{\ss}e 77, 50937 Cologne, Germany}
\author{Martin~Sundermann}
  \affiliation{Institute of Physics II, University of Cologne, Z{\"u}lpicher Stra{\ss}e 77, 50937 Cologne, Germany}
  \affiliation{Max Planck Institute for Chemical Physics of Solids, N{\"o}thnitzer Stra{\ss}e 40, 01187 Dresden, Germany}
\author{Fabio~Strigari}
  \affiliation{Institute of Physics II, University of Cologne, Z{\"u}lpicher Stra{\ss}e 77, 50937 Cologne, Germany}
\author{Jo~Kawabata}
	\affiliation{Department of Quantum Matter, AdSM, Hiroshima University, Higashi-Hiroshima 739-8530, Japan}
\author{Toshiro~Takabatake}
	\affiliation{Department of Quantum Matter, AdSM, Hiroshima University, Higashi-Hiroshima 739-8530, Japan}
\author{Arata~Tanaka}
  \affiliation{Department of Quantum Matter, AdSM, Hiroshima University, Higashi-Hiroshima 739-8530, Japan}
\author{Peter~Bencok}
\affiliation{Diamond Light Source, Didcot OX11 0DE, United Kingdom}
\author{Fadi Choueikani}
  \affiliation{Synchrotron SOLEIL, L'Orme des Merisiers, Saint-Aubin, BP 48, 91192 Gif-sur-Yvette C$\acute{e}$dex, France}
\author{Andrea~Severing}
 	\email{severing@ph2.uni-koeln.de}
	\affiliation{Institute of Physics II, University of Cologne, Z{\"u}lpicher Stra{\ss}e 77, 50937 Cologne, Germany}
	\affiliation{Max Planck Institute for Chemical Physics of Solids, N{\"o}thnitzer Stra{\ss}e 40, 01187 Dresden, Germany}

\date{\today}

\begin{abstract}
Here we present linear and circular polarized soft x-ray absorption spectroscopy (XAS) data at the Ce $M_{4,5}$ edges of the electron (Ir) and hole-doped (Re) Kondo semiconductor CeOs$_2$Al$_{10}$. Both substitutions have a strong impact on the unusual high N$\acute{e}$el temperature, $T_N$=28.5\,K, and also the direction of the ordered moment in case of Ir. The substitution dependence of the linear dichroism is weak thus validating the crystal-field description of CeOs$_2$Al$_{10}$ being representative for the Re and Ir substituted compounds. The impact of electron- and hole-doping on the hybridization between conduction and 4$f$ electrons is related to the amount of $f^0$ in the ground state and reduction of x-ray magnetic circular dichroism. A relationship of $cf$-hybridization strength and enhanced $T_N$ is discussed.  The direction and doping dependence of the circular dichroism strongly supports the idea of strong Kondo screening along the crystallographic $a$ direction. 
\end{abstract}


\maketitle

\section{introduction}
The ternary Ce compounds Ce$T$$_2$Al$_{10}$ ($T$=Fe, Ru and Os) crystallize in the orthorhombic YbFe$_2$Al$_{10}$-type structure\,\cite{Thiede1998, Tursina2005} and have experienced some attention due to the combination of Kondo semiconducting ground states and long range antiferromagnetic order (AF) in CeRu$_2$Al$_{10}$ and CeOs$_2$Al$_{10}$. Only the $T$\,=Fe compound remains paramagnetic (PM) down to 50\,mK.\cite{Strydom2009, Muro2009, Muro2010a, Muro2010b,Takesaka2010,Adroja2013} The presence of a magnetic order in these cage-like compounds is indeed surprising because the Ce-Ce distances of 5.2\AA are very large,\cite{Strydom2009,Nishioka2009} and even more unusual are the high ordering temperatures of $T_N$\,=\,27\,K (Ru) and 28.5\,K (Os). They are at odds with de-Gennes scaling that implies $T_N$ should be lower than 1\,K.\cite{Adroja2013} The magnetic properties are very anisotropic with the $a$ axis being the easy axis ($\chi_a$\,$\gg$\,$\chi_c$\,$\textgreater$\,$\chi_b$). Inelastic neutron scattering has shown that the ordered magnetic moments $\mu_{ord}$ are not aligned along the easy axis; instead they are parallel to the $c$ direction. The general magnetic anisotropy and the small ordered magnetic moments of about 0.4\,$\mu_B$ can be explained to a large extend with crystal-electric field (CEF) effects. However, the low temperature susceptibility $\chi_a$ is strongly reduced with respect to a CEF-only calculation.\cite{Yutani2012,Adroja2013,Strigari2012,Strigari2013} Spin flip transitions to $\mu_{ord}$\,$\|$\,$b$ have been observed for magnetic fields of 4-6\,T along the $c$ direction,\cite{Tanida2012,Kawabata2014} i.e. it is not easy to orient the ordered moment along the easy axis. All these oddities put into question whether conventional Ruderman-Kittel-Kasyua-Yosida (RKKY) interaction is the only driving mechanism for the formation of magnetic order. 
\begin{figure}[H]
	\centering
	 \includegraphics[width=0.98\columnwidth]{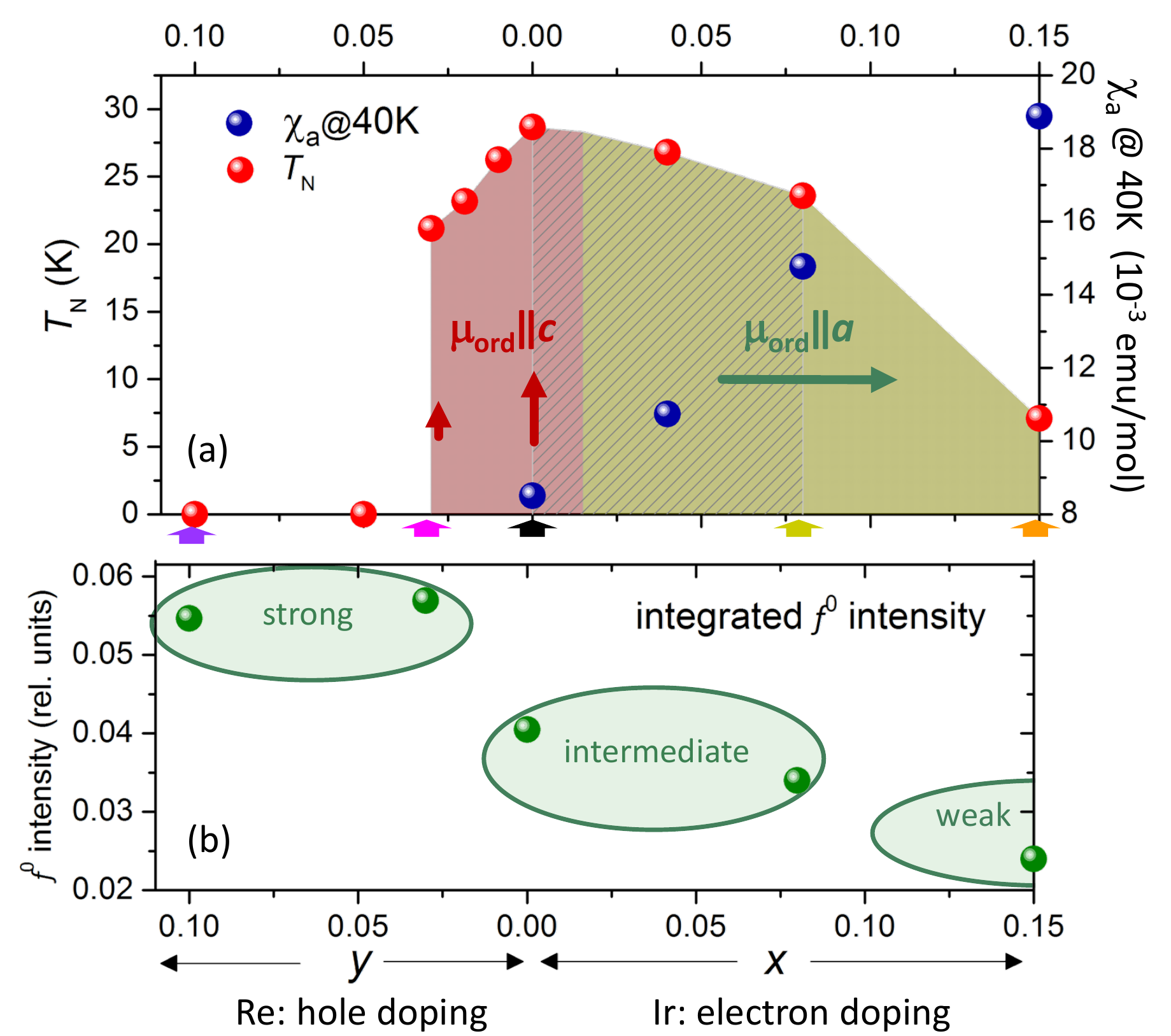}
		\caption{(a) Temperature\,-\,substitution phase diagram of Ce(Os$_{1-x}$Ir$_x$)$_2$Al$_{10}$ and Ce(Os$_{1-y}$Re$_y$)$_2$Al$_{10}$ showing $T_N$ (red dots) and the magnetic susceptibility at 40\,K (blue dots). Data are adapted from Ref.\,\onlinecite{Kawabata2014}. The red and green regions refer to different magnetically (AF) ordered phases, the length of the red and green arrows symbolizes the size of ordered moment. (b) Integrated $f^0$ intensities obtained from Fig.\,3\,(c) and (d) in relative units as function of substitution. The labels \textsl{strong}, \textsl{interim} and \textsl{weak} refer to the anticipated $cf$ hybridization strength of the respective substitution range.}
		\label{fig1}
\end{figure}

Like in many heavy fermion and Kondo materials, the hybridization of the $f$ electrons with the conduction band ($cf$ hybridization) plays an important role in the ground state formation of the Ce$T$$_2$Al$_{10}$ compounds.\cite{Adroja2013,Ishiga2014,Strigari2015} In Kondo lattice or heavy fermion compounds, the $cf$ hybridization leads to the opening of a hybridization gap. In so-called Kondo insulators or semiconductors the chemical potential lies in the gap so that at low temperatures non-metallic behavior is observed.\cite{Nishioka2009,Muro2009,Lue2010a,Muro2010a,Kawamura2010_Fe,Chen_NMR2010} Single crystalline resistivity data yield activation energies of the order of 4-8\,meV depending on the crystallographic direction. Excitation gaps have also been reported from the spin excitation spectra\,\cite{Adroja2013} and optical conductivity spectra.\cite{Kimura2011a,Kimura2011b,Kimura2011c} The anisotropies in the optical conductivity as well as very detailed lattice parameter measurements of the RE$T_2$Al$_{10}$ (RE = rare earth)\,\cite{Sera2013} suggest the presence of anisotropic hybridization with Kondo screening along the $a$ direction being most effective. It has been speculated that this may have an impact on the magnetic order.\,\cite{Kikuchi2014,Kikuchi2017} The $T$\,=\,Fe compound is the most strongly hybridized one among the family, thus explaining the absence of magnetic order. We suggest Ref.\,\onlinecite{Adroja2013} and references therein for a more detailed overview of the properties of the undoped Ce$T_2$Al$_{10}$ compounds.

\begin{figure}[b]
	\centering
	 \includegraphics[width=0.99\columnwidth]{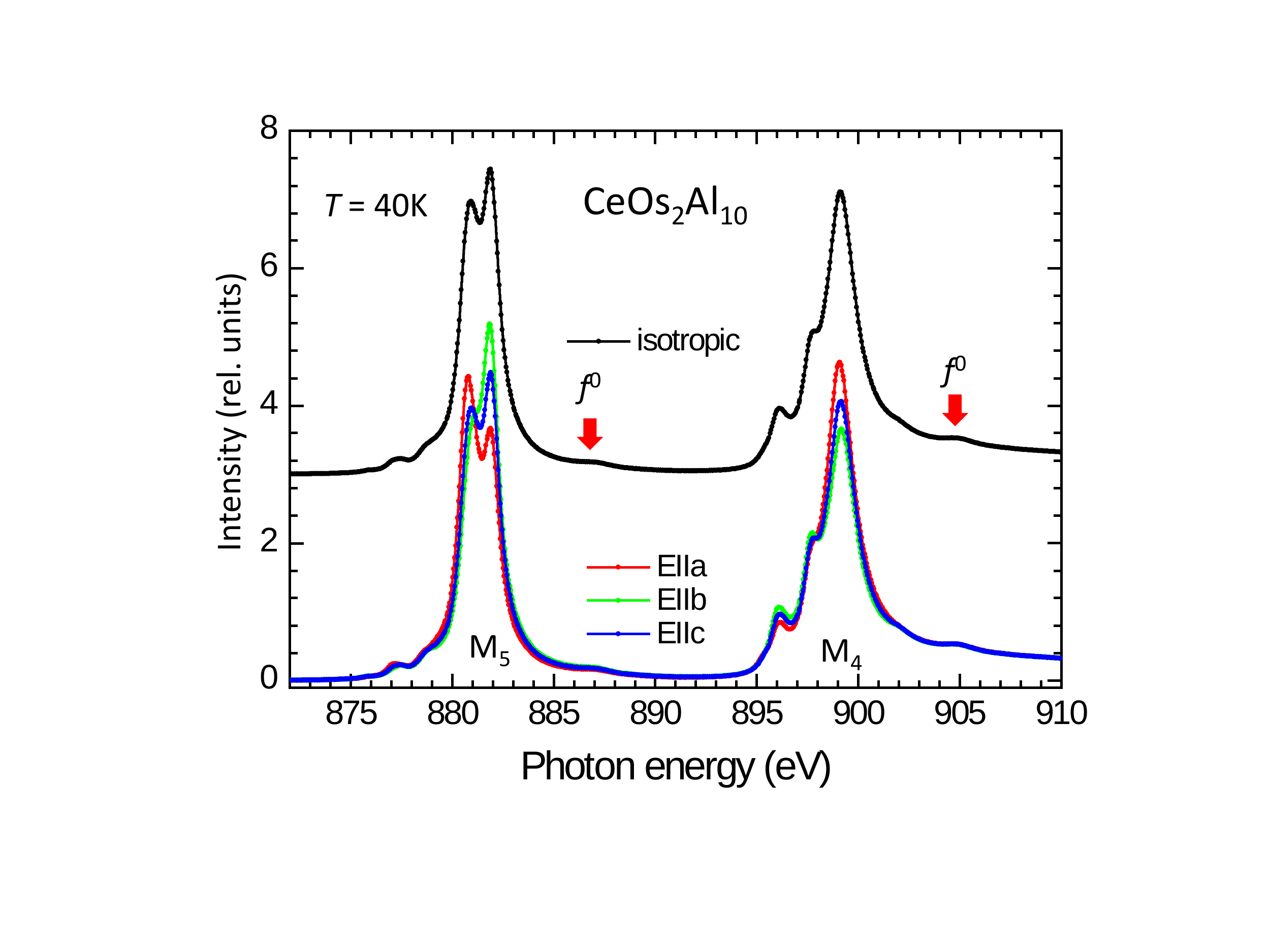}
		\caption{Experimental linear polarized M$_{5,4}$-edge XAS data of CeOs$_2$Al$_{10}$ at 40K: isotropic spectra I$_{iso}$\,=\,1/3(I$_{E\|a}$\,+\,I$_{E\|b}$\,+\,I$_{E\|c}$) (top, black curve) and polarized data with E$\|a$ (red), E$\|b$ (green) and E$\|c$ (blue). Red arrows point to enlarged regions shown in insets.}
		\label{fig2}
\end{figure}

The magnetic and transport properties of the Ce$T$$_2$Al$_{10}$ are very susceptible to doping\,\cite{Bhattacharyya2014a,Kimura2015,Adroja2015, Hayashi2016, Khalyavin2013,Kawabata2014, Bhattacharyya2014b, Khalyavin2014,Yamada2015, Kawabata2017} so that this family is an ideal playground for investigating itinerant versus localized magnetism in a Kondo system. The electron or hole doping with Ir or Re on the transition metal site in Ce(Os$_{1-x}$Ir$_{x}$)$_2$Al$_{10}$ or Ce(Os$_{1-y}$Re$_{y}$)$_2$Al$_{10}$ has a big influence on the magnetic order and other physical properties:\cite{Khalyavin2013,Kawabata2014,Bhattacharyya2014b,Khalyavin2014,Yamada2015,Kawabata2017} The low temperature static susceptibility $\chi_a$ increases dramatically with increasing Ir concentration (see blue dots in Fig.\,\ref{fig1}) while $\chi_{b,c}$ changes only little for the two other directions. For 15\% Ir the anisotropy of $\chi$ is well described within a local CEF model. Furthermore, the direction of the ordered magnetic moment flips from $c$ to the easy axis $a$ somewhere for $x$ between 0 and 0.03 and its size increases to 1\,$\mu_B$ for $x$\,=\,0.15 (see red and green regions and arrows in Fig.\,\ref{fig1}).\cite{Khalyavin2013,Kawabata2014} The N$\acute{e}$el temperature, however, decreases at first slowly from 28.5 to 20\,K for $x$ between 0 and 0.08 and then further to 7\,K for $x$\,=\,0.15 (see red dots in Fig.\,\ref{fig1}). This is contrasted by substitution with Re in Ce(Os$_{1-y}$Re$_{y}$)$_2$Al$_{10}$. The Re substitution drives the system into a more strongly hybridized regime so that T$_N$ and also the size of the ordered moment decrease. The direction of the ordered moment remains unchanged. Beyond $y$\,=\,0.05 magnetic order is suppressed. Also other compounds, e.g. CeRh$_3$B$_2$\,\cite{Dhar1981} and CeRh$_2$Si$_2$\,\cite{Quezel1984} exhibit unusual high ordering temperatures in the presence of $cf$ hybridization. All this suggests strongly that the presence of $cf$-hybridization is entangled with the enhancement of the N$\acute{e}e$l temperature.\cite{Hoshino2013} This aspect will be studied in a systematic manner at the example of CeOs$_2$Al$_{10}$ with Re and Ir substitutions.

Here we present Ce M$_{4,5}$ edge ($3d^{10}4f^1$\,$\rightarrow$\,$3d^{9}4f^2$) x-ray absorption spectroscopy (XAS) study of electron (Ir) and hole doped (Re) CeOs$_2$Al$_{10}$. The concentrations were chosen such that the phase diagram in Fig.\,\ref{fig1} is covered (see thick arrows); the undoped compound CeOs$_2$Al$_{10}$, Ir: $x$\,=\,0.08 and 0.15 and Re: $y$\,=\,0.03 and 0.1. Linear dichroism yields information about the doping dependence of the CEF wave functions and, thereby, information about the $cf$-hybridization can be obtained from the isotropic spectra that are constructed from the linear polarized data. The $f^0$ configuration contributes to the ground state in the presence of $cf$-hybridization and gives rise to small, extra spectral weights due to the transition $3d^{10}4f^0$\,$\rightarrow$\,$3d^{10}4f^1$. Their relative intensities allow sorting the members of the substitution series according to their hybridization strength. The magnetic circular dichroism (MCD), finally, gives information about the doping dependence of the magnetic moments, i.e. about the direction dependence of the Kondo screening. 

\section{Methods and Experimental Details}
XAS is an element specific spectroscopic method that is extremely sensitive to the valence, spin and orbital or crystal-field state of the ion under investigation.\,\cite{deGroot,TanakaJPSC63,footnote}  Applied to the excitation from a $3d^{10}4f^1$ CEF ground state to the final state $3d^{10}4f^2$ multiplets, the linear dichroism (LD) of XAS provides information about the CEF \textsl{ground state} wave function of the Ce $f^1$ configuration when measuring at such low temperatures that only the ground state is occupied.\cite{Hansmann2008} M-edge XAS is not resolution limited, because the excited CEF states only enter via thermal population, in contrast to inelastic neutron scattering,\cite{Adroja2013} (surface sensitive) high resolution angle resolved photoelectron spectroscopy (ARPES),\cite{Patil2014,Chen2018a} or M-edge resonant inelastic x-ray scattering (RIXS)\,\cite{Amorese2016} where the CEF transitions are directly probed. In the present study only the 4$f$ ground state is targeted and for this purpose M-edge XAS has shown to be most sensitive to small changes in the wave function upon substitution.\cite{Willers2015,Chen2018}

XAS at the Ce L-edge ($2p$\,$\rightarrow$\,$5d$) has shown to be very powerful for determining the 4$f$ occupation i.e. the amount of $f^1$ versus $f^0$ in the ground state in the presence of the core hole. The core hole effect due to M-edge absorption also leads to a large energy splitting of the final states involving the $f^1$ and $f^0$ configurations, so that we can use our data for giving a relative change of the amount of $f^0$ with substitution. Here we construct the isotropic spectra from the sum over all directions of the electric field vector $E$ of the linear polarized light.

The difference of the absorption of left or right circular polarized light (CD) depends on the magnetic properties of the ion under investigation. Here a magnetic field is applied in order to align the Ce moments. The resulting magnetic circular dichroism (MCD) is directly proportional to the magnetic moments aligned along the applied magnetic field.

Single crystals of CeOs$_2$Al$_{10}$,  $\rm Ce(Os_{1-x}Ir_x)_2Al_{10}$ ($x\leq0.15$) and $\rm Ce(Os_{1-y}Re_y)_2Al_{10}$ ($y\leq0.10$) were grown by an Al self-flux method \cite{Takesaka2010, Khalyavin2014,Kawabata2014} and their quality and orientation were confirmed by Laue x-ray diffraction. 

The x-ray linear dichroism (XLD) measurements were carried out on beamline BL10 at the synchrotron light source Diamond in the UK. The x-ray magnetic circular dichroism (XMCD) measurements were carried out on DEIMOS beamline at Synchrotron Soleil in France. The energy resolution at the Ce $M_{4,5}$ edge ($h\nu\approx870-910$\,eV) was about 0.15\,eV. The samples were inserted into the respective cryomagnets and cleaved \textsl{in situ} in the ultra high vacuum chambers with a pressure of $\sim\,10^{-10}$\,mbar. All spectra were recorded in the total-electron yield (TEY) by measuring the drain current of the sample.

Two differently oriented samples of each composition have been measured with horizontal and vertical linear polarized light so that linear polarized spectra could be taken along the three orthorhombic axis, i.e. for $E$\,$\|$\,$a$, $E$\,$\|$\,$b$, and $E$\,$\|$\,$c$. The linear dichroism (LD) is defined as the difference between a linear polarized spectrum I$_{E\|i}$, i\,=\,$a,b,c$ and the isotropic spectrum I$_{iso}$, LD$_i$\,=\,I$_{E\|i}$\,-\,I$_{iso}$ with I$_{iso}$\,=\,1/3(I$_{E\|a}$\,+\,I$_{E\|b}$\,+\,I$_{E\|c}$).

The XMCD data were measured with an applied magnetic field $B$\,=\,6\,T along the Poynting vector of the light. Right $\sigma_i^{+}$ and left polarized $\sigma_i^{-}$ XAS data were taken on three differently oriented crystals so that for each composition $B$\,$\|$\,$i$, $i$\,=\,$a,b,c$, could be realized. The magnetic circular dichroism (MCD) is defined as difference between the two circular polarizations, MCD$_i$\,=\,$\sigma_i^{+}$\,-\,$\sigma_i^{-}$.

For better comparison all data are normalized to the integrated intensity. The data simulation has been performed with the full multiplet code XTLS-830\,\cite{TanakaJPSC63} as described previously.\cite{Hansmann2008,Strigari2012,Strigari2013,Chen2016}

\section{Results}
The bottom spectra in Fig.\,\ref{fig2} are the linear polarized data of all three orthorhombic direction, I$_{E\|a}$, I$_{E\|b}$ and I$_{E\|c}$, of CeOs$_2$Al$_{10}$ at 40\,K. The data show a strong direction dependence, in agreement with data published in Ref.\,\onlinecite{Strigari2013}. 

The top spectrum in Fig.\,\ref{fig2} is an isotropic spectrum, constructed as introduced above. It exhibits, in addition to the main absorption lines due to the transition $3d^{10}4f^1$\,$\rightarrow$\,$3d^{10}4f^2$, some smaller humps that are due to the amount of $f^0$ in the ground state of these hybridized compounds. The red arrows point out the energy regions where the $f^0$ humps can be seen best. 

\begin{figure}
	\centering
	 \includegraphics[width=0.99\columnwidth]{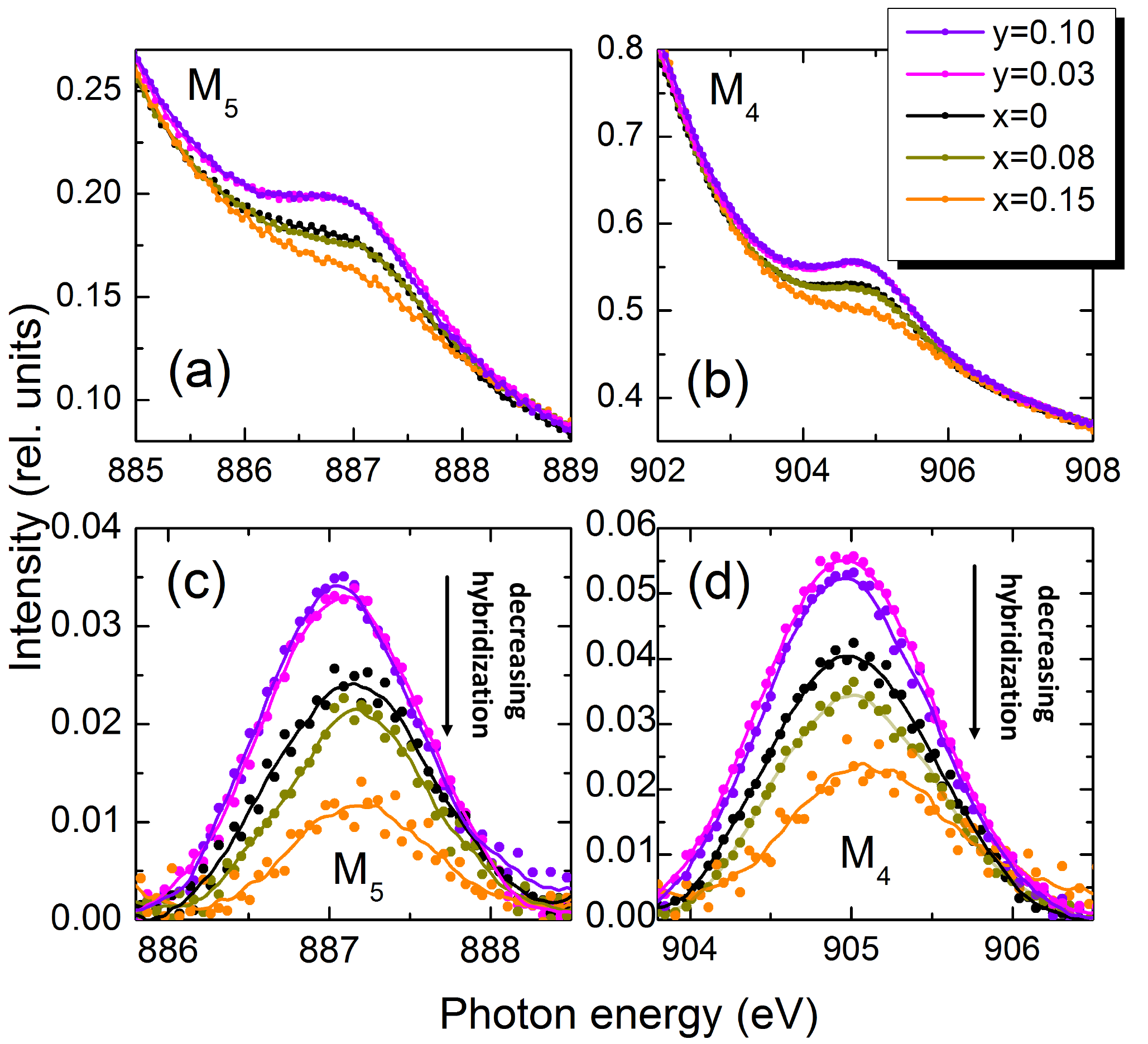}
		\caption{(a) and (b) $f^0$ regions at the Ce M$_{4,5}$ edges (see red arrows in Fig.\,2) of isotropic data for CeOs$_2$Al$_{10}$ and substitutions with Re and Ir. (c) and (d) Same $f^0$ regions as in (a) and (b) after subtracting a linear background.}
		\label{fig_new}
\end{figure}

Figure\,\ref{fig_new}\,(a) and (b) show the $f^0$ regions at the M$_{4,5}$ edges for all the measured samples on an enlarged scale. Figure\,\ref{fig_new}\,(c) and (d) display the same $f^0$ intensities but after subtracting a linear background. The vertical scales have the same relative intensities as in Fig.\,\ref{fig2}, i.e. the total of $f^0$ is small but clearly changes upon substitution. The integrated intensities, averaged over both edges, are displayed in Fig.\,\ref{fig1}\,(b) as function of substitution. Note, these numbers represent a relative and not the absolute $f^0$  occupation in contrast to the numbers given in Ref.\,\onlinecite{Strigari2015}. Nevertheless, The relative $f^0$ occupations tell us that there are three regions; the amount of $f^0$ is almost the identical for the pure ($x$\,=\,0) and $x$\,=\,0.08 Ir doped compounds, it is larger in the two hole doped Re compositions ($y$\,=\,0.03 and 0.1) and smallest for the highest electron doping (Ir, $x$\,=\,0.15). Although there is no one to one scaling of $f$-occupation and hybridization $V$ it is valid to say that there is the general trend in cerium compounds that a lager $f^0$ occupation goes along with stronger hybridization.\cite{Khomskii} Hence we find three regimes, a strongly hybridized one for the two Re doped samples, a region of interim hybridization ($x$\,=0 and 0.03) and a weakly hybridized one for $x$\,=\,0.15. 

\begin{figure}
	\centering
	 \includegraphics[width=0.99\columnwidth]{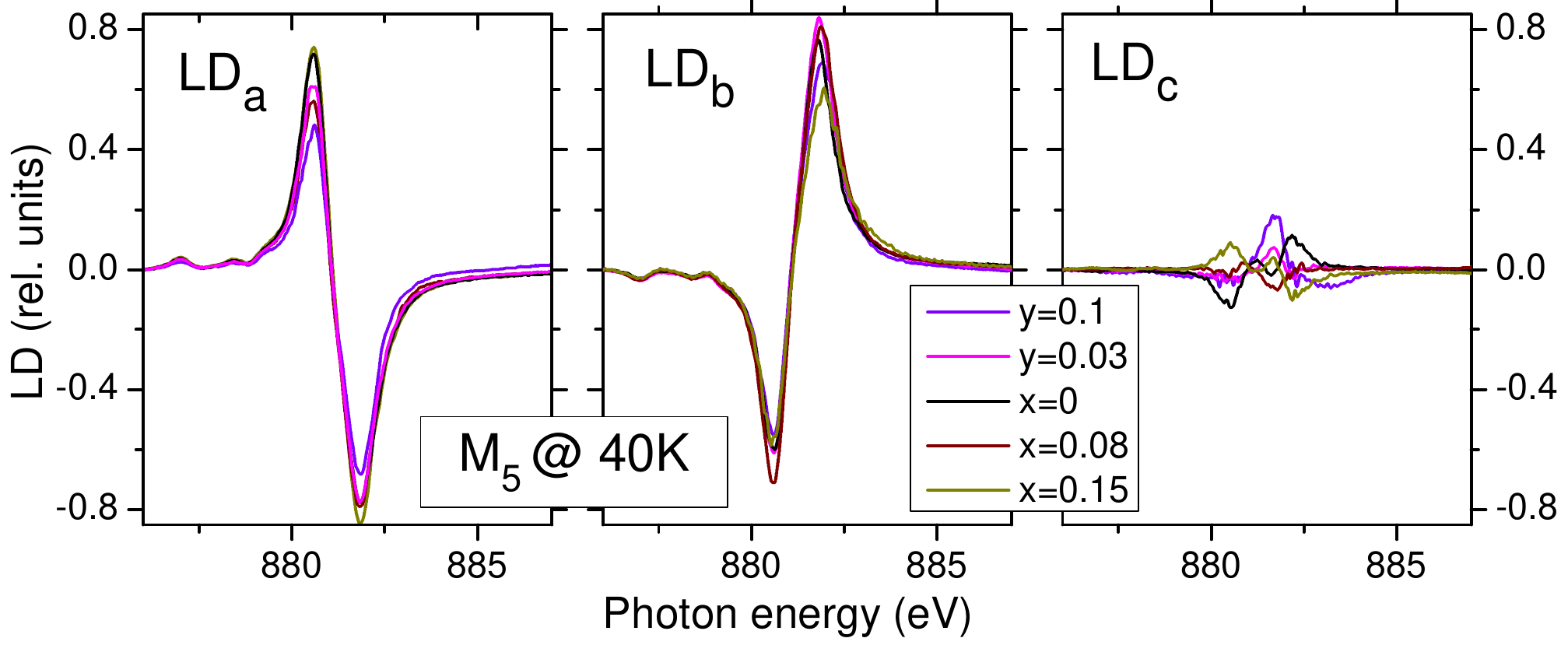}
		\caption{(a) Experimental linear dichroism LD$_i$ of Ce M$_5$-edge of all samples for the three crystallographic directions, i\,=\,$a,b,c$, with LD$_i$\,=\,I$_{E\|i}$\,-\,I$_{iso}$ and I$_{iso}$\,=\,1/3(I$_{E\|a}$\,+\,I$_{E\|b}$\,+\,I$_{E\|c}$). By definition $\sum(LD_i)$\,=\,0}
		\label{fig3}
\end{figure}

Figure\,\ref{fig3}\, shows the LD at the M$_5$ edge of all compositions for all three directions. There is some scatter in the LD when going from Ir to Re substitution but, in contrast to previous findings in substitution series of the Ce115 compounds,\,\cite{Willers2015,Chen2018} we cannot judge whether there is any systematic in the present data. We should recall that in these orthorhombic compounds all three directions contribute to LD$_i$ due to the isotropic spectra whereas we look at the difference of two directions in the tetragonal systems. Nevertheless, we draw the important conclusion that the CEF potential is only marginally affected by substitution and cannot be responsible for the dramatic increase of $\chi_a$ with electron doping (see blue dots in Fig.\,\ref{fig1}\,(a)).

\begin{figure}
	\centering
	 \includegraphics[width=0.99\columnwidth]{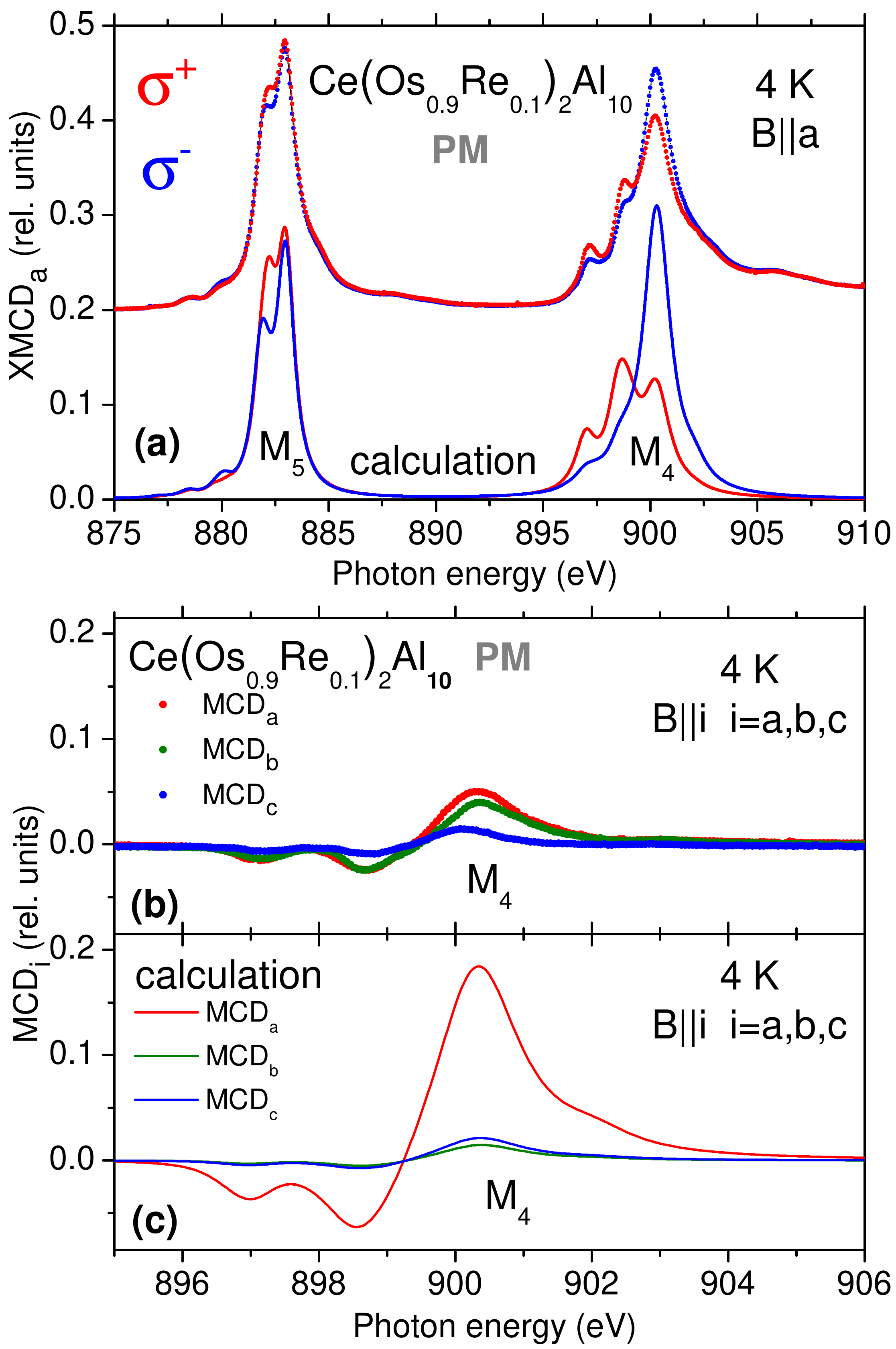}
		\caption{(a) Experimental circular polarized $M_{4,5}$-edge XAS data, $\sigma^+_a$ and $\sigma^-_a$, for $B$\,=\,6T$\|a$ of (a) paramagnetic, hole doped Ce(Os$_{0.9}$Re$_{0.1}$)$_2$Al$_{10}$ and the CEF-only simulation based on the Strigari \textsl{et al.} paramaters.\cite{Strigari2013} (b) Experimental and (c) calculated (CEF-only) MCD$_i$ with $i$\,=\,$a$, $b$, $c$ of Ce(Os$_{0.9}$Re$_{0.1}$)$_2$Al$_{10}$.}
		\label{fig4}
\end{figure}

\begin{figure}
	\centering
	 \includegraphics[width=0.99\columnwidth]{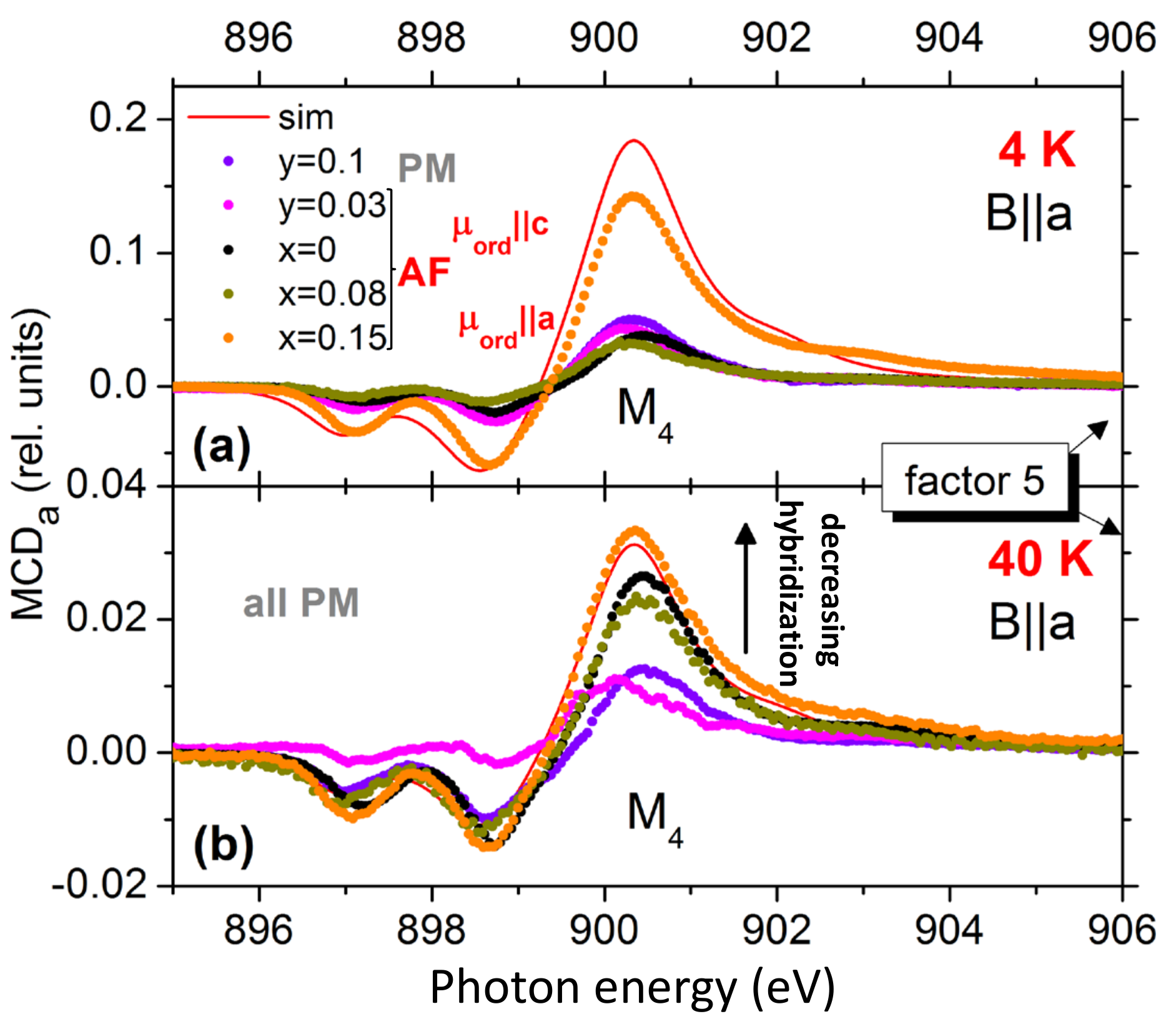}
		\caption{Circular dicroism MCD$_a$ of hole doped (Re, y=0.1 and 0.03), pure and electron doped (Ir, $x$\,=\,0.08 and 0.15) CeOs$_2$Al$_{10}$ (colored dots), compared to a CEF-only calculation\,\cite{Strigari2013} (red line) at (a) 4\,K and (b) in the paramagnmetic state at 40\,K .}
		\label{fig5}
\end{figure}

At 4\,K the MCD is much larger than at 40\,K (compare Fig.\,\ref{fig5}\,(a) and (b)). However at 4\,K all samples with the exception of Ce(Os$_{0.9}$Re$_{0.1}$)$_2$Al$_{10}$ are magnetically ordered.  We therefore show the XMCD data for $B$\,$\|$\,$a$ of this 10\% hole doped sample in Fig.\,\ref{fig4}\,(a), the direction dependence of its XMCD in Fig.\,\ref{fig4}\,(b) and the comparison to a CEF-only calculation (see bottom spectra in Fig.\,\ref{fig4}\,(a) and Fig.\,\ref{fig4}\,(c)). Evidently, the discrepancy between data and simulation is largest for MCD$_a$. We interpret this as a strong indication for the Kondo effect being most effective along the $a$-direction. 

Figure\,\ref{fig5}\,(a) and (b), show the measured MCD$_a$ for all substitutions (dots) at 4\,K and 40\,K, and the respective CEF-only simulations (red lines) for comparison. Note, that the y-axis in Fig.\,\ref{fig5}\,(d) is spread by a factor of 5 because the MCD-signal at 40\,K is much smaller with respect to 4\,K. At 4\,K all samples with the exception of $y$\,=\,0.10 are magnetically ordered so that the interpretation of these 4\,K data is not straight forward. The moments may be canted with respect to the field direction and no statement can be made about the possibility of Kondo screening. At 40\,K, however, all samples are paramagnetic. Here we observe, similar to the $f^0$ intensity, three regimes: strongly suppressed moments for $y$\,=\,0.1 and 0.03, almost recovered moments for $x$\,=\,0 and 0.08 and a fully recovered moment for $x$\,=0.15. This coincides nicely with substitution dependence of the $f^0$ intensity in Fig.\,\ref{fig_new}, thus supporting the labeling of the samples as strongly, intermin and weakly hybridized as in Fig\,\ref{fig1}\,(b).

\section{discussion}
X-ray absorption has been used to investigate the relative amount of $f^0$, the LD and MCD at 6\,T in several hole and electron doped CeOs$_2$Al$_{10}$ samples. The linear polarized data show that the CEF description of CeOs$_2$Al$_{10}$ is a good Ansatz for describing the anisotropy of the substitution series. It verifies that the doping dependence of the magnetic susceptibility as in Ref.\,\onlinecite{Kawabata2014} is not due to changes in the CEF.  The analysis of the isotropic data that were constructed from the linear polarized ones give relative occupation numbers of the $f^0$ configuration in the Ce(Os$_{1-y}$Re$_{y}$)$_2$Al$_{10}$ and Ce(Os$_{1-x}$Ir$_{x}$)$_2$Al$_{10}$ compounds. The $f^0$ occupation as well as the MCD$_a$ intensity show the same trend as function of substitution (see black arrows in Fig.\,\ref{fig_new}\,(c) and (d) and Fig.\,\ref{fig5}\,(b). There are three regimes of hybridization: Ce(Os$_{1-y}$Re$_{y}$)$_2$Al$_{10}$ for $y$\,=\,0.1 and 0.03 being strongly hybridized, CeOs$_2$Al$_{10}$ and Ce(Os$_{0.92}$Ir$_{0.08}$)$_2$Al$_{10}$ exhibit interim $cf$ hybridization and Ce(Os$_{0.85}$Ir$_{0.15}$)$_2$Al$_{10}$ being weakly hybridized.
 
Interestingly, the variation of $T_N$ also suggest three regimes (see Fig.\,\ref{fig1}): The region of suppressed magnetic order, i.e. $T_N$\,=\,0 that coincides with strong $cf$-hybridization, the region of strongly enhanced $T_N$ according to de Gennes scaling, from $y$\,=0.03 to $x$\,=\,0.08 where $cf$-hybridization is interim (see dashed region in Fig.\,\ref{fig1}\,(a)). Within this dashed region the ordered moments are at first aligned along $c$, but already for an Ir concentration of $x$\,=\,0.03 the ordered moment is aligned along the easy axis $a$,\cite{Bhattacharyya2014a} suggesting a weakening of the Kondo screening along $a$.  Beyond $x$\,=\,0.08, $T_N$ decreases to 7\,K and the $cf$ hybridization is smallest. This suggests strongly that the presence of of interim $cf$-hybridization drives the enhancement of the N$\acute{e}e$l temperature or, following Hoshino and Kuramoto, the magnetic response is strongly enhanced at the itinerant to localized transition.\cite{Hoshino2013}  

A uniaxial pressure study of CeRu$_2$Al$_{10}$ and CeOs$_2$Al$_{10}$ shows that also other factors contribute to enhancing $T_N$.\cite{Hayashi2017} Hayashi \textsl{et al.} found that uniaxial pressure along the $b$ axis enhances $T_N$ without increasing hybridization. It was concluded that charge conduction along the $b$ axis plays an important role in the ordered phase. Here should be kept in mind, as pointed out in Ref.\,\cite{Hayashi2017}, that the shortest Ce-Ce distance is along $b$ and that uniaxial pressure in $b$ direction  reduces this distance. Our finding that $cf$ hybridization is one of the driving mechanisms in enhancing $T_N$ in the CeOs$_2$Al$_{10}$ compounds with Re and Ir substitution does not contradict this pressure study.   

Finally, the detailed analysis of the XMCD data strongly suggests that the $a$ direction is most affected by Kondo screening. In particular, the most strongly hybridized compound Ce(Os$_{0.9}$Re$_{0.1}$)$_2$Al$_{10}$ that is paramagnetic at all temperatures has a strongly reduced MCD$_a$ signal with respect to the CEF-only simulation (see Fig.\,\ref{fig4}\,(b),\,(c)) This is contrasted by Ce(Os$_{0.85}$Ir$_{0.15}$)$_2$Al$_{10}$ which is the least hybridized sample. Here the MCD$_a$ signal has completely recovered above the ordering temperature (see Fig.\,\ref{fig5}\,(b)) and is almost recovered due to a spin flip transition in the antiferromagnetic state with $\mu_{ord}$$\|$$a$ (see Fig.\,\ref{fig5}\,(a)). 

\section{Summary}
Ce(Os$_{1-y}$Re$_y$)$_2$Al$_{10}$ ($y$\,=0.03 and 0.1) and Ce(Os$_{1-x}$Ir$_x$)$_2$Al$_{10}$ ($x$\,=\,0, 0.08, and 0.15) have been studied with linear and circular polarized x-ray absorption at the Ce $M_{4,5}$-edge. We find that the crystal-field description of the undoped compound describes the linear dichroism of all substitutions, i.e. is a good approximation of the CEF-only magnetic anisotropy. The MCD along the crystallographic $a$ direction is most affected by substitution. Here anisotropic Kondo screening is a likely explanation. XLD and MCD data, both show three regimes of hybridization, the strongest one for the two Re substitutions, an interim one for $x$\,=0 and 0.08 and the smallest for $x$\,=\,0.15. The strongest enhancement of the N$\acute{e}$el temperature $T_N$ coincides with the region of interim hybridization.

\section{Acknowledgement} K.C., M.S. and A.S. benefited from financial support from the Deutsche Forschungsgemeinschaft (DE), SE1441-1-2. T.T acknowledges the financial support from JSPS (JP), JP26400363 and JP16H01076. We acknowledge Synchrotron-Soleil for provision of synchrotron radiation facilities. Part of this work was carried out with the support of the Diamond Light Source (proposal SI11558).

\end{document}